\documentclass[showpacs,prb,superscriptaddress,twocolumn]{revtex4}
\usepackage{textcomp}
\usepackage{amsmath}
\usepackage{amssymb}
\usepackage{graphicx}

\usepackage{color}
\usepackage[normalem]{ulem}

\newcommand{\ket}[1]{|#1\rangle}
\newcommand{\bra}[1]{\langle #1|}
\newcommand{\ud}{\mathrm{d}}

\newcommand{\Tr}{\textrm{Tr}}

\newcommand{\mean}[1]{\langle #1\rangle}

\newcommand{\hH}{\hat{H}}

\newcommand{\corr}[1]{\mean{\hat{X}(#1)\,\hat{X}}}

\newcommand{\ww}[3]{i\langle #1(#3)\ket{\dot{#2}(#3)}}

\begin{document}

\title{Floquet theory of Cooper pair pumping}

\author{Angelo Russomanno}
\affiliation{NEST, Scuola Normale Superiore and Istituto Nanoscienze -- CNR,  56126
Pisa, Italy}

\author{Stefano Pugnetti}
\affiliation{NEST, Scuola Normale Superiore and Istituto Nanoscienze -- CNR, 56126
Pisa, Italy}

\author{Valentina Brosco}
\affiliation{ISC -- CNR and Dipartimento di Fisica, Universit\'a La Sapienza, P.le A. Moro 2, 00185 Roma, Italy}

\author{Rosario Fazio}
\affiliation{NEST, Scuola Normale Superiore and Istituto Nanoscienze -- CNR, 56126
Pisa, Italy}

\begin{abstract}
In this work we derive a general  formula  for the charge pumped in a superconducting nanocircuit.  Our expression generalizes
previous results in several ways, it is applicable both in the adiabatic and in the non-adiabatic regimes and it takes into
account also the effect of an external environment.  More specifically, by applying
Floquet theory to Cooper pair pumping, we show that under a cyclic evolution the
total charge transferred through the circuit is proportional to the derivative of the associated Floquet quasi-energy with
respect to the superconducting phase difference.   In the presence of an external environment the expression for the
transferred charge acquires a transparent form in the Floquet representation. It is given by the weighted sum of the charge
transferred in each Floquet state, the weights being the diagonal components of the stationary density matrix of the
system expressed in the Floquet basis. In order to test the power of this formulation we apply it to the study of pumping
in a Cooper pair sluice.  We reproduce the known results in the adiabatic regime and we show new data in the non-adiabatic
case.

\end{abstract}


\maketitle

\section{Introduction}
In a mesoscopic conductor a dc charge current can be obtained, in the absence of applied voltages, by cycling in time two
(or more) external parameters, e.g. gate voltages and/or magnetic fluxes, which govern the transport properties of the
system~\cite{thouless83}.  Adiabatic charge pumping refers to the regime when the variation of the
external parameters is slow as compared to the characteristic time scale of the system.

In the scattering approach to quantum transport the pumped charge in an adiabatic cycle can be expressed in terms of derivatives of the scattering
amplitudes with respect to the pumping parameters~\cite{brouwer98}.  In the opposite regime of Coulomb blockade, with
several metallic islands connected to each other by small tunnel junctions, a  periodic modulation of the externally applied gate-voltages
leads to a periodic lifting of Coulomb blockade, thus enabling the transfer of exactly one electron per period through the device.
Experimental evidence for parametric charge pumping in normal metallic systems in the regime of Coulomb blockade has
been obtained in Refs.~\onlinecite{kouwenhoven91,pothier92}.  Over the last decades charge pumping has attracted the interest of
many research groups working on very different aspects of this phenomenon ranging from its metrological applications to its
intimate relation with the fundaments of quantum theory (see Ref.~\onlinecite{pumpgeneral} and references therein).

Originally motivated by the aim of achieving quantized charge pumping in the GHz range, a great deal of attention has been devoted
in the last two decades to superconducting systems. The first experiment in this context, performed by Geerligs {\em et al}~\cite{geerligs91},
showed that the degree of quantization of the pumped charge was not as good as in the normal case. As it was later discussed
by Pekola {\em et al} ~\cite{pekola99}, the main source of inaccuracies is related to the overall coherence of the superconducting system.
This apparent disadvantage (appropriate designs of the superconducting circuit may overcome this difficulty) however turned out to
be a precious source for the investigation of fundamental properties of quantum theory in macroscopic systems.

If only superconducting leads are present, at low enough temperature, pumping is due to the adiabatic transport of Cooper pairs.
Besides the dependence of the pumped charge on the details of the cycle, in superconducting pumps there is an additional dependence
on the superconducting phase difference since the overall process is coherent. Cooper pair pumping has been thoroughly
investigated~\cite{pekola99,pekola01,aunola03,fazio03,niskanen03,governale05,mottonen06,leone08,brosco08,pirkkalainen10} in the last decades.
In a series of experiments the Helsinki group~\cite{vartianen07,mottonen08}  has shown the coherent properties of Cooper pair pumping
and, very importantly, provided the first experimental demonstration of the relation between Cooper pair pumping and the Berry phase
acquired by the system during its cyclic evolution.  A connection between Berry phase and pumped charge in superconducting nano-circuits
has been already established theoretically  in Refs.~\onlinecite{aunola03,governale05,mottonen06} (see also 
Refs.~\onlinecite{avron00,zhou03} where this relation was found for mesoscopic normal conductors).

Berry phases in macroscopic systems such as superconducting circuits have been studied in Refs.~\onlinecite{falci00,wang02,blais03,mottonen06} and
very recently experimentally demonstrated by the ETH group~\cite{leek07}. The large body of theoretical understanding and
the spectacular experimental control which lead to unveil the coherent properties of pumping and its relation to geometric
phases are important steps towards the implementation of geometric quantum computation~\cite{jones00,zanardi99} with superconducting
devices.

While a lot has been found in the relation between geometric phases and pumping in closed quantum systems, the role of an
external environment constitutes, with the notable exception in few very recent papers~\cite{pekola10,solinas10,salmilehto10},  an almost
unexplored territory.  The study of geometric phases in the presence of decoherence and dissipation has started only recently with
few exceptions though, certainly prompted by the interest in quantum computation (see for example Ref.~\onlinecite{geodiss}). 
Together with many features common in the theory of open
quantum systems, the analysis of decoherence in geometric interferometry rises several distinct issues that are of interest both as
fundamental questions in quantum mechanics and in quantum computation. The adiabatic evolution, for example, cannot occur
arbitrarily slow, as decoherence would destroy any interference. This implies that the decoherence processes should be analyzed in
close connection with non-adiabatic corrections~\cite{wubs06}.

Being Cooper pair pumping a geometric quantum effect, it is natural to ask oneself to which extent an external environment modifies its
characteristics.  Not only, this question is relevant for a detailed comparison with experimental
data where an external bath is unavoidably present, but also it may
shed additional light in the role of dissipation on geometric quantum phenomena. It is not a priori obvious, for example, that a relation
between pumping and Berry phases (of any sort), will survive in an open system. As already mentioned, till now this problem was
tackled in Refs.~\onlinecite{pekola10,solinas10,salmilehto10}, where a generalized master equation to consistently account for the combined action of the driving
and dissipation was derived.  Application to the Cooper pair sluice~\cite{mottonen08} showed that in the zero temperature limit the
ground state dynamics, and consequently pumping,  is not affected by the environment.

Stimulated by the results obtained in Refs.~\onlinecite{pekola10,solinas10,salmilehto10}, in this paper we further investigate the relation between pumping and
geometric phases. We derive an expression for the pumped charge which generalizes previous results in several
aspects. It is valid also under non-adiabatic conditions and in the presence of an external environment. The key to our
approach is to apply Floquet theory  to Cooper pair pumping. We will show that under cyclic evolution of the system the total charge
transferred through the circuit is proportional to the derivative of the associated Floquet quasi-energy with  respect to the superconducting
phase difference (a result which is valid also in the case of a non-adiabatic evolution).  In the presence of an external environment the
expression for the transferred charge is easily generalized in the Floquet representation. It is given by the weighted sum of the charge
transferred in each Floquet state, the weights being the diagonal components of the stationary density matrix of the system expressed in
the Floquet basis. The central result of our work is Eq.~(\ref{qcicnp53}); it embraces all the limits considered so far in the literature and allows
to investigate new regimes. Furthermore it suggests the use of a series of well known numerical schemes to compute the pumped charge.

The paper is organized as follows.  In the Sections~\ref{cppsection} and~\ref{floquet} we will introduce the basic ingredients needed in the derivation
of the pumping formula. In Section~\ref{cppsection} we formulate the problem of Cooper pair pumping in superconducting circuits while in
Section~\ref{floquet} we provide the necessary tools of the Floquet theory of driven quantum systems both in the closed and open cases.
The formula for the pumped charge will be derived in Section~\ref{floquetpump}.  Here we will discuss in which aspects our results
generalizes previous works.  As an example we will apply our derivation to the Cooper pair sluice which was experimentally
realized in the Helsinki group. In Section~\ref{sluicesection} various different limits will be discussed.  Section~\ref{conclusions} will contain
the conclusions  of the present work.

\section{Cooper pair pumping}
\label{cppsection}

Charge pumping in Josephson networks consists in a coherent periodic manipulation
of the collective state of the Cooper pairs in the array. In this Section we define the setting and
review, for later convenience, the relation between Cooper pair pumping and Berry phases.

\begin{figure}[h]
\begin{center}
\includegraphics[width=
\columnwidth]{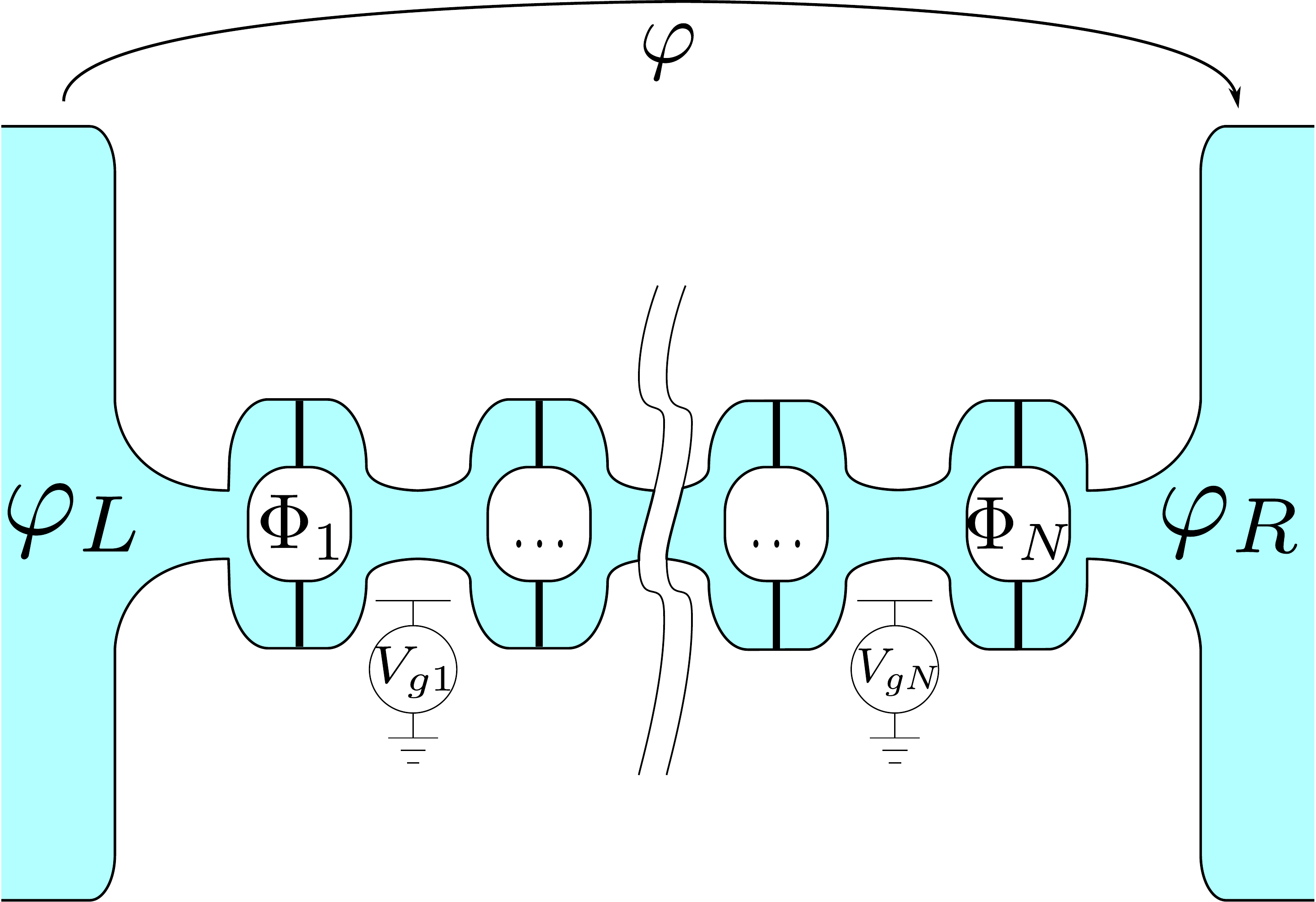}
\end{center}
\caption{A generic setup of a Cooper-pair pump. Two superconducting leads, kept at a phase
bias $\varphi = \varphi_R - \varphi_L$, are connected through a Josephson network. The system is
operated in a regime where quantum effects are important. In order to have a pumped charge
some external parameters (e.g. gate voltages or magnetic fluxes) are varied periodically. }
\label{generalsetup}
\end{figure}

A Cooper pair pump consists of a Josephson network connected through Josephson
junctions to two superconducting leads (see a sketch of the setup in Fig.\ref{generalsetup}).
The system is phase biased, i.e.~the two superconducting
electrodes are kept at a finite phase difference $\varphi = \varphi_R - \varphi_L$ where
$\varphi_{R/L}$ is the phase of the  superconducting order parameter of the right/left lead.
The Cooper pair pump operates by changing adiabatically in time some external parameters such
as gate voltages, to tune the charging energies, or magnetic fluxes,  to vary the effective Josephson
couplings. We will label this set of external parameters by  the vector $\vec {\lambda}(t)=\{V_{gi}(t),\Phi_{i}(t)\}$.
In the absence of an external environment charge transport is a purely coherent phenomenon.
The Hamiltonian of the pump depends on the superconducting phases of each island of the network
$\varphi_i$ ($i=1,\ldots N$), its conjugate momenta (i.e.~the charge on each island $n_i$), the phase difference across the
pump and all the external parameters, $H(t)=H\big[\varphi_1,...,\varphi_N;n_1,...,n_N;\vec{\lambda}(t), \varphi\big]$.
The state of the system is denoted by $|\Psi(t)\rangle= |\Psi (t,\vec{\lambda}(t),\varphi)\rangle$.

%%%%%%%%%%%%%%%%%%%%%%%%%%%%%%%%%%%%%%%%%%%%%%
%%%% F I G U R E   1 %%%%%%%%%%%%%%%%%%%%%%%%%
%%%%%%%%%%%%%%%%%%%%%%%%%%%%%%%%%%%%%%%%%%%%%%

By changing the control parameters in time a charge $Q^{(tr)}$ will be transferred from  the left to the right electrode.
The total charge transferred through the pump during the period $T$ is given by:
\begin{equation}
	Q^{(tr)}=-2e\dfrac 1{\hbar}\int_0^T \langle \Psi(t') |\frac{\partial H}{\partial \varphi }|\Psi(t')\rangle dt \;\; .
\label{qtr}
\end{equation}
Assuming that there are no degeneracies in the spectrum and in the adiabatic limit,  it was shown~\cite{aunola03,mottonen06}
that $Q^{(tr)}$ can be expressed in terms of the total phase accumulated by the system after the cycle
\begin{equation}
	\frac{Q^{(tr)}}{2e}=  \frac{\partial \gamma_D}{\partial \varphi } + \frac{\partial \gamma_B}{\partial \varphi } .
\label{pumpedberry}
\end{equation}
In the previous expression $\gamma_{D/B}$ are the dynamical/geometric contribution to the accumulated phase. The dynamical
contribution is the charge transferred through the circuit due to the supercurrent flow. The geometric contribution is the
pumped charge.  In the latter case the charge is an even function of the superconducting phase difference while the contribution
due to the supercurrent flow is odd in $\varphi$. The different symmetry under reflection of the phase bias is fundamental for the
experimental detection of the pumped charge.
A generalization to the non-abelian case, i.e.~in the presence of degeneracies in the spectrum, has been
given in Ref.~\onlinecite{brosco08}.

In order to investigate non-adiabatic corrections and to generalize this result to the dissipative case, it is useful to re-express
the pumped charge using the Floquet formalism.  In the next sections, after introducing the basic definitions of Floquet theory, we
will find an expression for the pumped charge which in the adiabatic limit and in the absence of an external environment reduces
to Eq.~(\ref{pumpedberry})

\section{Floquet theory}
\label{floquet}

As it will become clear in the continuation of the paper,  Floquet formalism is naturally suited to study Cooper pair pumping. It is
important to stress already now that this is not a mere re-formulation of what has been done so far. We will
show that, on the contrary, Floquet approach treating on the same footing adiabatic and non-adiabatic regimes, provides a transparent and general expression for the pumped charge in the case in which the superconducting network is
coupled to an external environment.  A Floquet scattering theory has been developed by  Moskalets and B\"{u}ttiker to study
pumping in mesoscopic conductors~\cite{moskalets02-04}. Here we employ Floquet approach to study Cooper pair pumping.

In the next sections we introduce the necessary ingredients of Floquet theory and its use in quantum dissipative systems. The presentation
follows closely Refs.~\onlinecite{grifoni98,guerin93}.

\subsection{Basics of Floquet theory}
\label{basicsfloquet}

Given a system whose dynamics is governed by a periodic Hamiltonian $\hat{H}(t)=\hat{H}(t+T)$,
Floquet theorem states that solutions to the Schr\"odinger equation exist which have the (Floquet) form
\begin{equation}
	\ket{\Psi_\alpha(t)}=e^{-i\epsilon_\alpha t/\hbar}\ket{\Phi_\alpha(t)}\;\;,
\label{floform}
\end{equation}
where the state $\ket{\Phi_\alpha(t)}$ is called Floquet mode and it is periodic in time ($\ket{\Phi_\alpha(t+T)}=\ket{\Phi_\alpha(t)} $) and the corresponding quasi-energy  $\epsilon_\alpha$ is real and  unique up to multiples of $\hbar\Omega$, with $\Omega=2\pi/T$.
There are as many distinct such solutions as the dimension of the Hilbert space $\mathcal{H}$. These solutions are linearly independent
and form a basis of the Hilbert space. An eigenvalue equation for the quasi-energy $\epsilon_\alpha$ can be obtained by defining the operator
$ \overline{H}(t)\equiv \hat{H}(t)-i\hbar\partial_t $
\begin{equation}
\label{lord}
	\overline{H}(t)\ket{\Phi_\alpha(t)}=\epsilon_\alpha\ket{\Phi_\alpha(t)} \;\; .
\end{equation}
The Floquet modes
$\ket{\Phi_{\alpha,n}(t)}=\ket{\Phi_{\alpha}(t)}\exp(-in\Omega t)$
with integer $n$ lead to a solution identical to the one given in Eq.(\ref{floform}), but with shifted quasi-energy
$ \epsilon_\alpha\rightarrow\epsilon_{\alpha,n}=\epsilon_\alpha-n\hbar\Omega $;
hence the eigenvalues $\lbrace\epsilon_\alpha\rbrace$ can be mapped in a first Brillouin zone obeying to $-\hbar\Omega/2\leq\epsilon\leq\hbar\Omega/2$.

For the Hermitian operator $\overline{H}(t)$ it is convenient to introduce the composite Hilbert space~\cite{sambe73} $\mathcal{H}\otimes\mathcal{T}$ made
by the tensor product of the Hilbert space $\mathcal{H}$ of the vectors representing the state of the system and the space $\mathcal{T}$ of
the periodic functions in $t$ with period $T=2\pi/\Omega$.  In the space of the periodic functions of $t$ we have a basis of Fourier
vectors $\lbrace\exp(-in\Omega t)\rbrace$ which are orthonormal with respect to the scalar product given by
\begin{equation}
	\label{ogdu}
	(m,n)\equiv\frac{1}{T}\int_0^T(e^{-im\Omega t})^*e^{-in\Omega t}\ud t=\delta_{m,\,n}
\end{equation}
We define $\ket{l}\equiv e^{-il\Omega t}$ as vectors in $\mathcal{T}$. We can extend the scalar product on $\mathcal{H}$ to a scalar product on $\mathcal{H}\otimes\mathcal{T}$ defining
\begin{equation}
	\langle\bra{\Psi_1}\Psi_2\rangle\rangle\equiv\frac{1}{T}\int_0^T\langle\Psi_1(t)\ket{\Psi_2(t)}\ud t \;\;.
\end{equation}

\subsection{Floquet states and geometric phases}
The Floquet quasi-energies are intimately connected to geometric phases (in the present work we are interested only in the
case in which their spectrum is non-degenerate). Indeed they are, up to a multiplying factor, the phases of the eigenvalues of the
evolution operator $\hat{U}_S(t+T,t)$. The phase is defined up to $2n\pi$, hence an eigenvalue of $\hat{U}_S(t+T,t)$
corresponds to infinite Floquet exponents obtained through translations of $2n\pi\hbar/T$.

Noticing that Floquet states follow a cyclic evolution in the projective Hilbert space (we call $\hat{C}$ the closed path followed in
the projective Hilbert space), it is possible to express the Aharonov-Anandan geometric phase acquired during a cyclic evolution starting in the 
Floquet eigenstate $\alpha$ by
\begin{equation}
	\gamma_{AA}(\hat{C}) = -\frac{\epsilon_{\alpha}T}{\hbar}+\frac{1}{\hbar}\int_0^T\bra{\Phi_\alpha(t)}\hat{H}(t)\ket{\Phi_\alpha(t)}\ud t
\label{bart1}
\end{equation}
An equivalent expression, perhaps more useful in the computation is
\begin{equation}
	\label{bart3}
	\gamma_{AA}(\hat{C}) = 2\pi\sum_kk\bra{c_{\alpha,k}}c_{\alpha,k}\rangle \;\; .
\end{equation}
where we used the Fourier expansion of the Floquet modes
\begin{equation}
	\label{mofou}
	\ket{\Phi_{\alpha}(t)}=\sum_{l=-\infty}^\infty\ket{c_{\alpha,l}}e^{-il\Omega t} \;\; .
\end{equation}

In the adiabatic limit the quasi-energy corresponding to the $n$-th eigenstates can be expressed in terms of the dynamic, $\gamma_D$,
and geometric, $\gamma_G$, phases:
\begin{equation}
\label{flubber}
	\epsilon_n=-\frac{\hbar}{T}[\gamma_{D,n}(T)+\gamma_{G,n}(C)] \; \; .
\end{equation}

\subsection{Floquet-Born-Markov Master equation}
\label{mastersection}
The Floquet basis is particularly useful to write the Master equation governing the dynamics of the reduced density matrix of a
driven system when in contact with an external environment. We consider below the case in which the Born-Markov approximation
is applicable. Details and subtleties of the derivation of a Master  equation in this case are described in Ref.~\onlinecite{grifoni98}, here we merely state the
end result for the Master equation which will be later used to derive a formula for the pumped charge.

Given a quantum system interacting with an external reservoir, the Hamiltonian describing the
system+reservoir is given by $	\hH_{S+R}=\hH(t)+\hH_R+\hat{V}$,
where $\hH(t)$ is periodic with period $T$ and the interaction has the form $\hat{V}=\hat{X}\otimes\hat{Y}$, where the
operator $\hat{X}$ acts on the environment, and $\hat{Y}$ on the system.

The Master equation for the reduced density matrix of the system can be
presented in the form~\cite{grifoni98}
\begin{widetext}
\begin{equation}
\label{happ1}
		\dot{\rho}_{\,\alpha\beta}(t) = \sum_{\gamma\delta;k,k'}\big[\Gamma_{\delta\beta\alpha\gamma ,k'k}^-+\Gamma_ {\delta\beta\alpha\gamma,k'k}^+
		 -\sum_{\nu}(\delta_{\beta\delta}\Gamma_{\alpha\nu\nu\gamma,kk'}^++\delta_{\alpha\gamma}\Gamma_{\delta\nu\nu\beta,kk'}^-)\big]\rho_{\,\gamma\delta}(t)
		e^{i(\Delta_{\alpha\beta,k}-\Delta_{\gamma\delta,-k'})t}\;\;,
\end{equation}
\end{widetext}
having defined
\begin{equation}
\nonumber
	\label{gamma}
	\begin{array}{lll}
		\Gamma_{\alpha\beta\gamma\delta,kk'}^+& = & \frac{1}{\hbar^2}Y_{\alpha\beta,k}Y_{\gamma\delta,k'}\gamma_{\gamma\delta,k'}^+\\
		&&\\
		\Gamma_{\alpha\beta\gamma\delta,kk'}^-& = & \frac{1}{\hbar^2}Y_{\alpha\beta,k}Y_{\gamma\delta,k'}\gamma_{\alpha\beta,k}^-\\
		&&\\
		Y_{\alpha\beta,k}& = &\frac{1}{T}\int_0^{T}\bra{\Phi_\alpha(t)}\hat{Y}\ket{\Phi_\beta(t)}e^{i\Omega k t}dt \\
		&& \\
		\gamma_{\alpha\beta,k}^\pm & = & \int_0^\infty \corr{\pm t''}\exp(-i\Delta_{\alpha\beta,k}t'')\ud t'' \\
		&& \\
		\Delta_{\alpha\beta,k} &= &\frac{1}{\hbar}(\epsilon_\alpha-\epsilon_\beta)-k\Omega \;\;.
	\end{array}
\end{equation}
As it is evident from Eq.~(\ref{happ1}), Floquet theory allows one to treat the time periodic case with a formalism which is formally
identical to the one used in the time-independent case. The relevant effects due to the periodic driving are captured by the
use of the Floquet basis.

Further simplifications can be made if the secular approximation holds \cite{grifoni98}(as we will assume in the rest of the paper). As in the time-independent
case, the equations for the populations decouple from those for the (off-diagonal) coherences. In the steady state the coherences vanish.
The populations are given by a "detailed balance" condition
\begin{equation}
	W_{\nu\rightarrow\alpha}\rho_{\,\nu\nu}^{\textrm{st}}=W_{\alpha\rightarrow\nu}\rho_{\,\alpha\alpha}^\textrm{st}
	\label{pappst}
\end{equation}
with
\begin{equation}
	\label{exto1}
	W_{\delta\rightarrow\alpha}=\frac{1}{\hbar^2}\sum_{k}|Y_{\delta\alpha,k}|^2g(\Delta_{\delta\alpha,k})
\end{equation}
and
\begin{equation}
	\label{gugo}
	g(\omega)\equiv\int_{-\infty}^{+\infty}\corr{t}e^{i\omega t}\ud t \;\; .
\end{equation}
This property of the stationary solution, when expressed in the Floquet representation, is crucial to obtain the
pumped charge also in the dissipative case.

\section{Floquet approach to Cooper pair pumping}
\label{floquetpump}

Equipped with the Floquet formalism outlined above, we now derive an expression for the pumped charge both in the absence and in the
presence of an external environment. We first consider the case of a unitary evolution where, in the adiabatic case, we should recover the
known relation, Eq.~(\ref{pumpedberry}), between pumping and geometric phases.

\subsection{Pumped charge in a closed system}
At first we ignore any coupling with the external environment. The dynamics is unitary and governed by a time-periodic Hamiltonian.
It is meaningful to compute the pumped charge in a given cycle only for those states that, up to a phase, do come back to their initial value.
These are the Floquet states (one should keep in mind that no assumption of an adiabatic dynamics is done at this point).
By employing the Schr\"odinger equation in Eq.(\ref{qtr}) it is possible to write it in the form
\begin{equation}
	Q^{(tr)}=-2ie\int_0^T\partial_t   \langle \Psi(t) | \partial_{ \varphi} |\Psi(t)\rangle dt \;\; .
\label{qtr1}
\end{equation}
Since the average is performed over a Floquet state, defined in Eq.(\ref{floform}), it straightforward to obtain
\begin{equation}
	Q^{(tr)}=- 2e\frac{T}{\hbar}\partial_\varphi\epsilon_\alpha(\varphi) \;\; .
\label{qcicn8}
\end{equation}
Eq.(\ref{qcicn8}) is the first result of this paper. It gives a general formula for the transferred charge in a superconducting circuit which is valid
{\em both} under adiabatic and non-adiabatic conditions. Obviously it reduces to Eq.(\ref{pumpedberry}) in the adiabatic case.

The pumped contribution to the transferred charge can be obtained by subtracting from Eq.(\ref{qcicn8}) the supercurrent term (associated to
the dynamical phase)
\begin{equation}
	Q_p = 4\pi e\sum_kk\,\partial_\varphi\bra{c_{\alpha,k}(\varphi)}c_{\alpha,k}(\varphi)\rangle \;\; .
\label{qcicn9}
\end{equation}

\subsection{Pumped charge of a superconducting circuit coupled to an external environment}

The Floquet approach allows for a very simple and appealing generalization of Eq.(\ref{qcicn9}) to the dissipative case. For simplicity we consider
the case in which the secular approximation holds.
The transferred charge for a system defined by a density matrix is given by
\begin{equation}
	Q^{(tr)} = \frac{2e}{\hbar}\int_0^T\Tr\Big(\big(\partial_{\varphi}\hat{H}(t,\varphi)\big)\hat{\rho}^{\textrm{st}}(t,\varphi)\Big)\ud t
	\label{qcicn51}
\end{equation}
where $\hat{\rho}^{\textrm{st}}$ is the reduced density matrix of the system in the steady state (we are interested in
obtaining a suitable expression for the pumped charge in the stationary limit after all transient effects have disappeared). Noting that (see the previous Section) in the
Floquet basis all the coherences vanish in the long time limit, the expression for the transferred charge takes the form
\begin{equation}
	\label{qcicn53}
	Q^{(tr)}=-2e\frac{T}{\hbar}\sum_{\nu}\rho_{\nu\nu}^{\textrm{st}}\partial_\varphi\epsilon_\nu(\varphi)
\end{equation}
Hence, the charge passing through the circuit is the weighted average of the charge which would have passed if the system
had been in a pure Floquet state, see Eq.(\ref{qcicn8}). The weights are the populations of these Floquet states in the quasi-stationary case. The previous expression
can be split in a geometric and a dynamic part and the pumped charge is given by
\begin{equation}
	\label{qcicnp53}
	Q_p= 4\pi e\sum_{k,\nu}k\rho_{\,\nu\nu}^{\textrm{st}}(\varphi)\partial_\varphi\bra{c_{\nu,k}(\varphi)}c_{\nu,k}(\varphi)\rangle \;\; .
\end{equation}
This is the central result of our work. Eq.(\ref{qcicnp53}) reduces to all known cases in the corresponding limits. In addition allows us to
explore regimes that have not been considered so far.  The form in the dissipative case is self-explaining, it is the average of the corresponding
expression in the noiseless case weighted by the populations of the Floquet states.

In the rest of the paper we will apply the general result of Eq.(\ref{qcicnp53}) to the Cooper pair sluice~\cite{mottonen08} which, as we will
see in the next section, can be described by a two-level Hamiltonian. It is therefore useful to give explicit formulas in the case of a
two-dimensional Hilbert space.  In this case we have only two independent Floquet states that we call $\ket{\Psi_{\alpha}}$ and
$\ket{\Psi_{\beta}}$. The populations in the stationary state are given by
\begin{eqnarray}
	\label{qaspop}
	\rho_{\,\alpha\alpha}^{\textrm{st}}&=&\frac{W_{\beta\rightarrow\alpha}}{W_{\beta\rightarrow\alpha}+W_{\alpha\rightarrow\beta}}\nonumber\\
	\rho_{\,\beta\beta}^{\textrm{st}}&=&\frac{W_{\alpha\rightarrow\beta}}{W_{\beta\rightarrow\alpha}+W_{\alpha\rightarrow\beta}}
\end{eqnarray}
Replacing this in  Eq.(\ref{qcicnp53})  and exploiting the relation  $\epsilon_{\alpha}=-\epsilon_{\beta}$ we obtain
\begin{equation}
\label{qpump53}
	Q_p= 4\pi e\frac{W_{\beta\rightarrow\alpha}(\varphi)-W_{\beta\rightarrow\alpha}(\varphi)}{W_{\alpha\rightarrow\beta}(\varphi)+
	W_{\beta\rightarrow\alpha}(\varphi)}\sum_{k}k\partial_\varphi\bra{c_{\alpha,k}(\varphi)}c_{\alpha,k}(\varphi)\rangle.
\end{equation}

\section{Pumping in the Cooper pair sluice}
\label{sluicesection}
\begin{figure}
	\begin{center}
 	\includegraphics[width=\columnwidth]{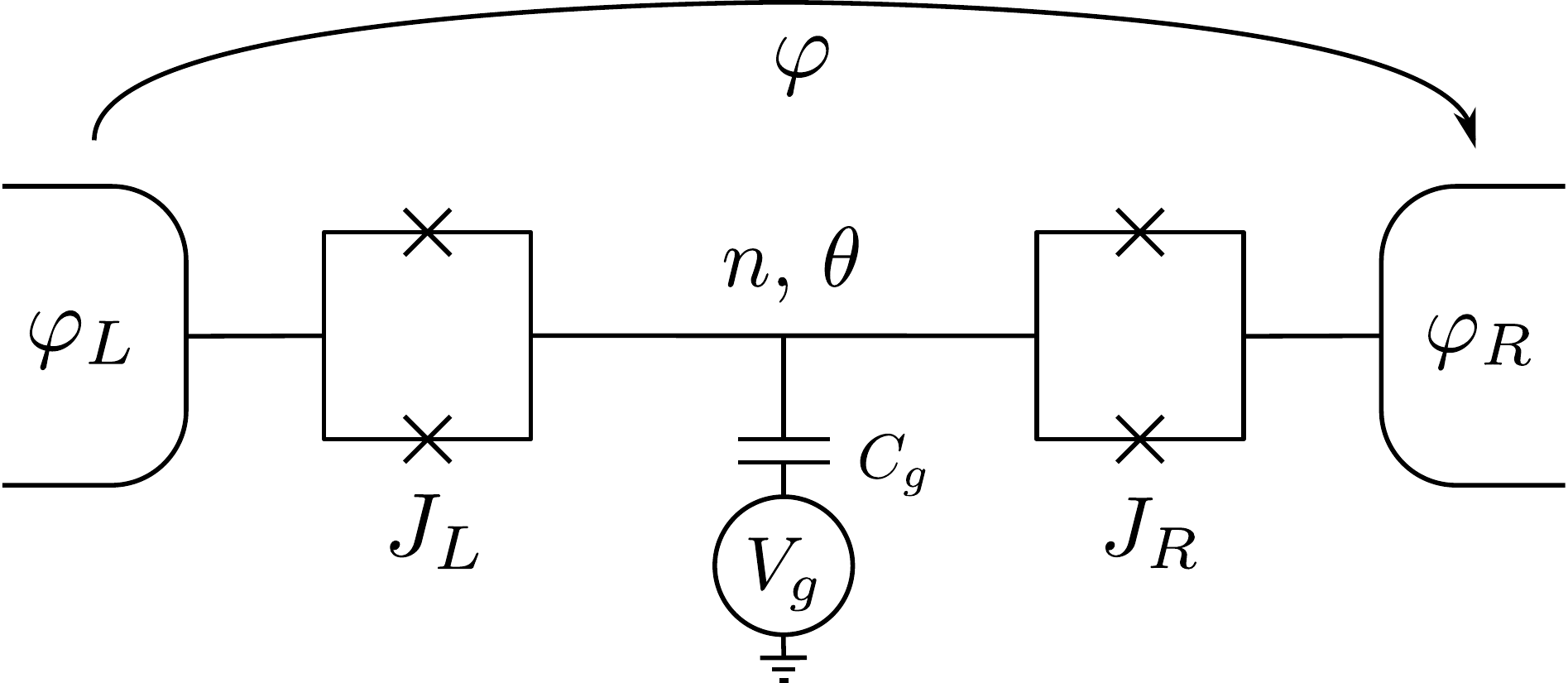}
	\end{center}
	\caption{Circuit scheme of the Cooper pair sluice.}
	\label{sluice}
\end{figure}

\begin{figure}
	\begin{center}
 	\includegraphics[width=\columnwidth]{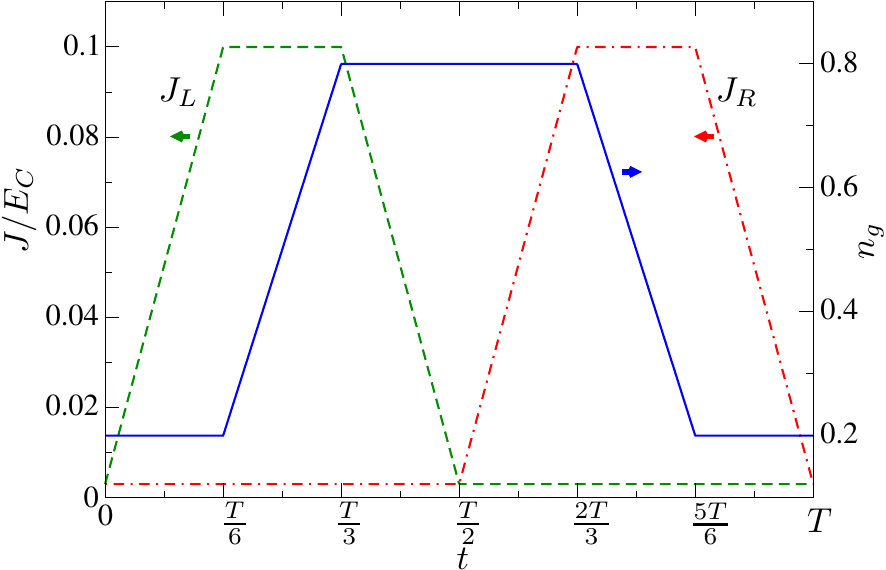}
	\end{center}
	\caption{The variation in a cycle of $n_g$, $J_L$ e $J_R$. These latter  are expressed in units of $E_C$. In this graph
	$n_g$ varies between $n_{g\textrm{ min}}=0.2$ and $n_{g\textrm{ max}}=0.8$, $J_L$ and $J_R$ vary between
	$J_{\textrm{min}}=0.003E_C$ and $J_{\textrm{max}}=0.1E_C$.}
	\label{pattu}
\end{figure}

The Cooper pair sluice~\cite{mottonen08} is a superconducting transistor where the pumping effect is achieved by a modulation
of the Josephson couplings and the gate voltages. A sketch of the sluice is shown in Fig.\ref{sluice}. The central island is connected to the
two superconducting leads by two tunable Josephson junctions. Its charging energy can be tuned by means of a gate voltage.
The Hamiltonian of the sluice is
\begin{equation}
\label{rhamy1}
	\hat{H} = E_C \big(\hat{n}-n_g\big)^2-\sum_{i=L/R} J_i\cos(\hat{\theta}-\varphi_i)
\end{equation}
where $E_C$ is the charging energy of the central island, $J_i$ are the Josephson coupling to the left ($i=L$) and right ($i=R$)
electrodes and $n_g$ is the gate charge which can be modulated by changing the gate voltage $V_g$ as shown in Fig.\ref{sluice}.
%%%%%%%%%%%%%%%%%%%%%%%%%%%%%%%%%%%%%%%%%%%%%%
%%%% F I G U R E   2 %%%%%%%%%%%%%%%%%%%%%%%%%
%%%%%%%%%%%%%%%%%%%%%%%%%%%%%%%%%%%%%%%%%%%%%%
In order to tune the Josephson couplings the junctions are replaced by SQUIDs which behave as single Josephson junctions
with an effective coupling which can be varied by changing the flux piercing the loop $J(\Phi)=J^{(0)}\cos (\pi \Phi / \Phi_0)$
(with $\Phi$ the flux through the loop and $\Phi_0$ the flux quantum). The charge $\hat{n}$ on the central island and the phase
$\hat{\theta}$ are canonically conjugated variables. The Hamiltonian is varied along a cyclic path, we change periodically
$\Phi_L$, $\Phi_R$ e $V_g$, determining a periodic variation of $n_g$, $J_L$ e $J_R$.
In all the cases considered here, these parameters are assumed to depend on time in the same way as in Pekola  {\em et al}~\cite{pekola10}.
The time-dependence of the parameters is shown in Fig.\ref{pattu}.
%%%%%%%%%%%%%%%%%%%%%%%%%%%%%%%%%%%%%%%%%%%%%%
%%%% F I G U R E   3 %%%%%%%%%%%%%%%%%%%%%%%%%
%%%%%%%%%%%%%%%%%%%%%%%%%%%%%%%%%%%%%%%%%%%%%%
Notice that when $J_L$ is maximum $J_R$ is minimum and vice versa, meaning that  when a SQUID is open the other
is closed and vice versa.  In an ideal situation where the minimum value of the Josephson couplings could be reduced to zero
the charge passing through the system in one cycle would be exactly quantized in units of $2e$ (the supercurrent contribution
vanishes in this setup). In a realistic case where the SQUID loops do not close perfectly the pumped charge is given
by~\cite{mottonen08}
\begin{equation}
\label{pekola}
	Q_p \sim -2e \Big(1-2\frac{J_{\textrm{min}}}{J_{\textrm{max}}}\cos\varphi \Big)
\end{equation}

If $n_g$ during the cycle stays close  enough to the degeneracy value $1/2$, we can describe the system using the two-level
Hamiltonian
\begin{equation}
\label{h23}
	H=-\frac{1}{2}(B_x\hat{\sigma}_x+B_y\hat{\sigma}_y+B_z\hat{\sigma}_z)
\end{equation}
where
$B_x=E_C(1-2n_g)$, $ B_y=J_R\sin\varphi$,  and $B_z=(J_L+J_R\cos\varphi)$.

In the following sections we will use the two-level approximation to present our results for the pumped charge in the Cooper pair
sluice.

\subsubsection{Unitary evolution}

\begin{figure}
	\begin{center}
	\includegraphics[width=\columnwidth]{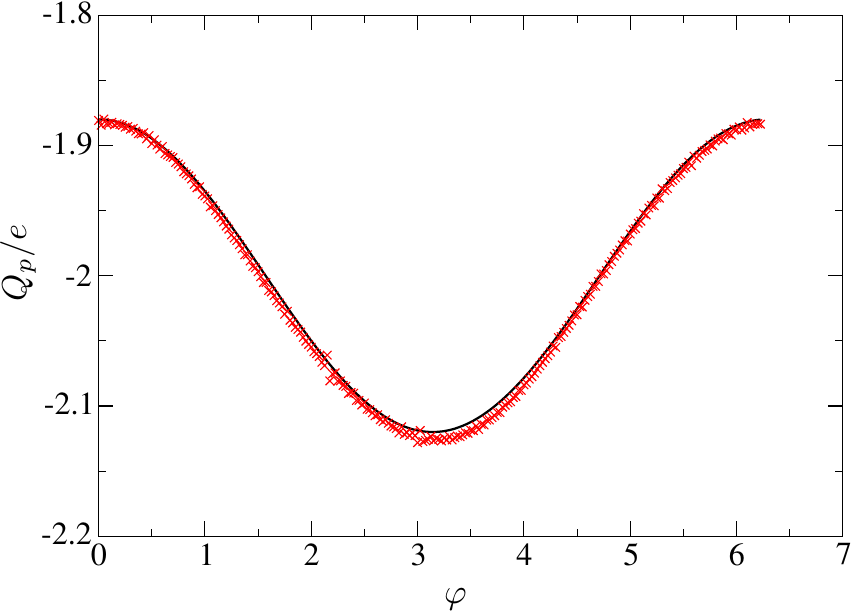}
	\end{center}
	\caption{Pumped charge vs.~$\varphi$ in the adiabatic case computed with the Floquet theory (red crosses) confronted with the same quantity computed
	with the analytical expression (\ref{pekola}) (black solid line), $0\leq\varphi\leq2\pi$. It is
	$J_\textrm{min}/J_\textrm{max}=0.03$, $J_\textrm{max}=0.1E_C$, $T E_C/\hbar=8400$. Charge is in units of $e$.}
	\label{fada1}
\end{figure}

We first consider the case in which the environment is absent. We compute numerically the Floquet exponents for the Hamiltonian
defined by  Eq.(\ref{h23})  and by means of Eq.(\ref{qpump53}) we obtain the pumped charge. The first case we consider is the
adiabatic limit to compare our approach with the known results. The behavior of the pumped charge as a function of the phase bias
is shown in Fig.\ref{fada1}.

%%%%%%%%%%%%%%%%%%%%%%%%%%%%%%%%%%%%%%%%%%%%%%
%%%% F I G U R E   4 %%%%%%%%%%%%%%%%%%%%%%%%%
%%%%%%%%%%%%%%%%%%%%%%%%%%%%%%%%%%%%%%%%%%%%%%

The results of the Floquet approach are tested against the analytic result~\cite{mottonen08}, Eq.(\ref{pekola}).
We chose  $T E_C/\hbar=8400$, a value which guarantees amply to be in the adiabatic limit. The reason for this large value is due only to
the simplicity to compute numerically the Floquet exponents. Obviously we do not expect any changes in the results as long as we are
in the adiabatic regime. The numerical calculations agree well with the analytical expression of Eq.(\ref{pekola}).

\begin{figure}
	\begin{center}
	\includegraphics[width=\columnwidth]{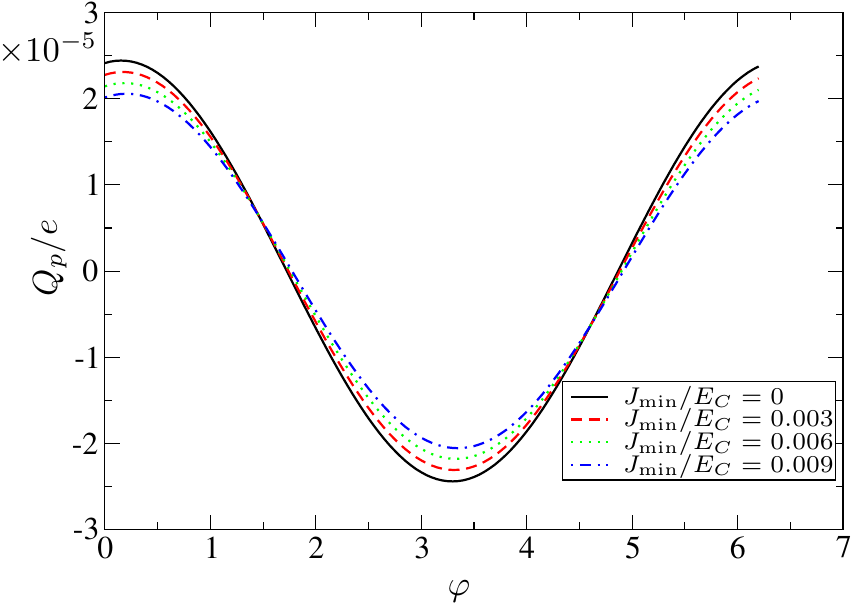}
	\end{center}
	\caption{Pumped charge vs $\varphi$ with $T E_C/\hbar=2.1$ for different values of $J_\textrm{min}$, $0\leq\varphi\leq2\pi$. Charge is in units of $10^{-5}e$.}
	\label{p21}
\end{figure}

As we already discussed, the Floquet approach to pumping allows us to go beyond the adiabatic regime.
An example is shown in Fig.\ref{p21} for $T E_C/\hbar=2.1$. In this particular case the pumped charge
is much smaller than that obtained in the adiabatic regime. This example was
indeed chosen just to demonstrate the power of the Floquet approach. There are cases, for a suitable choice of the parameters'  loop,
in which it is possible to obtain charge  quantization also under non-adiabatic conditions.

It is interesting to note that in the non-adiabatic case the pumped charge is phase-dependent also in the case in which 
$J_\textrm{min}=0$. This case is very similar to the Cooper pair shuttle~\cite{gorelik01,romito03}.

%%%%%%%%%%%%%%%%%%%%%%%%%%%%%%%%%%%%%%%%%%%%%%
%%%% F I G U R E   5 %%%%%%%%%%%%%%%%%%%%%%%%%
%%%%%%%%%%%%%%%%%%%%%%%%%%%%%%%%%%%%%%%%%%%%%%

\subsubsection{Influence of the external environment}

The second application of our pumping formula deals with the case in which the sluice is coupled to an external environment.
The main source of decoherence is due to charge fluctuations. In terms of the coupling Hamitonian introduced in~\ref{mastersection},
the operator of the system $\hat{Y} \propto \sigma_z$ and $\hat{X}=E_C\,\hat{\delta n_g}$,
where $\delta \hat{n}_g=C_g\hat{V}_g/2e$ expresses the fluctuations of the gate voltage ($C_g$ is the gate capacitance).
As usual, we assume that the charge fluctuations are due to the thermal noise of a resistance $R'$ put in series with $C_g$.
There are also some fluctuations in the fluxes $\Phi_L$ and $\Phi_R$, but these ones are coupled to the Josephson energies $J_L$ and
$J_R$ which are $J_L,J_R\ll E_C$, hence these fluctuations are much smaller than the charge fluctuations which are coupled to $E_C$
and we can neglect them.
The function $g(\omega)=\int_{-\infty}^{\infty}\corr{t''}e^{i\omega t''}\ud t''$ which appears in Eq.(\ref{exto1}) is defined as
\begin{equation}
	\label{omic}
	g(\omega)=\omega R\Big(e\frac{C_g}{C_\Sigma}\Big)^2 \Big(\coth\Big(\frac{\beta\hbar\omega}{2}\Big)+1\Big)
\end{equation}
In all the subsequent calculations $R=300\textrm{k}\Omega$.

\begin{figure}
	\begin{center}
	\includegraphics[width=\columnwidth]{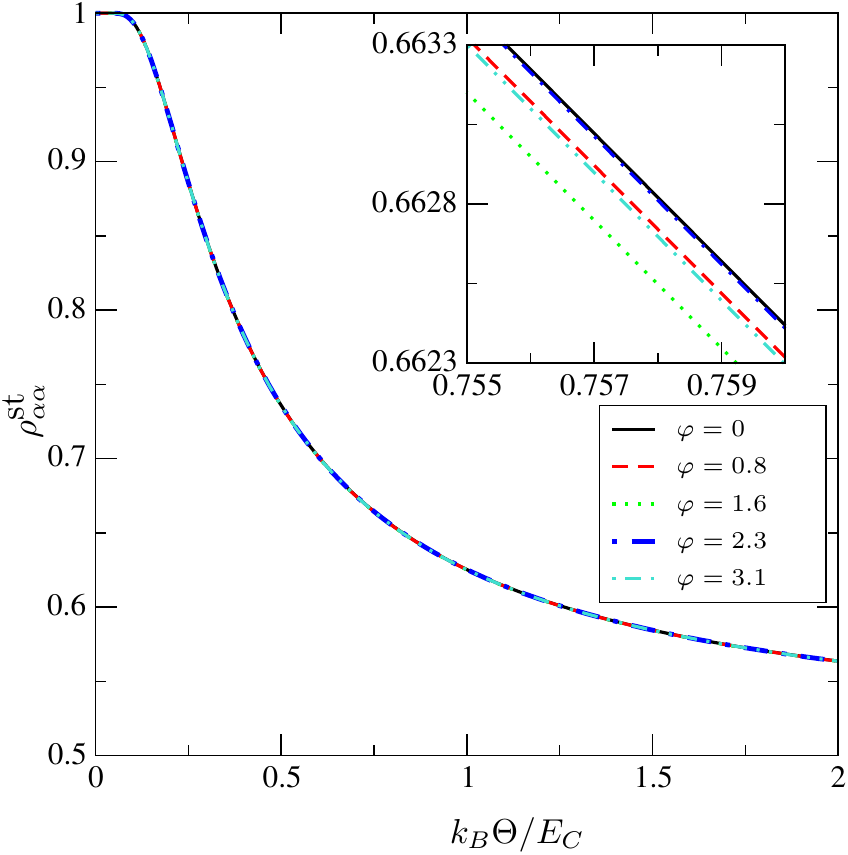}
	\end{center}
	\caption{Population $\rho^\textrm{st}_{\alpha\alpha}$ of the Floquet state with the lower quasi-energy vs temperature $\Theta$ (in its dimensionless form) for different
	values of $\varphi$. We are in the adiabatic limit ($T E_C/\hbar=8400$ and $\Delta E_{\textrm{min}}\simeq E_C/10$). Temperature
	is expressed in units of $E_C/k_B$. (Inset) The same graph magnified to show the dependence on $\varphi$.}
	\label{popul}
\end{figure}

There are several methods to compute the Floquet quasi-spectrum~\cite{grifoni98}.  We diagonalized numerically the Floquet
operator $\bar{H}(t)$ in the composite Hilbert space~\cite{sambe73} as briefly discussed in Section~\ref{basicsfloquet}. From the
knowledge of the Floquet eigenvalues and eigenvectors it is possible to compute the steady state populations in each Floquet mode.
These latter are shown in Fig.\ref{popul}. The parameters are chosen to be such that the pumping is adiabatic.

%%%%%%%%%%%%%%%%%%%%%%%%%%%%%%%%%%%%%%%%%%%%%%
%%%% F I G U R E   6 %%%%%%%%%%%%%%%%%%%%%%%%%
%%%%%%%%%%%%%%%%%%%%%%%%%%%%%%%%%%%%%%%%%%%%%%

\begin{figure}
	\begin{center}
	\includegraphics[width=\columnwidth]{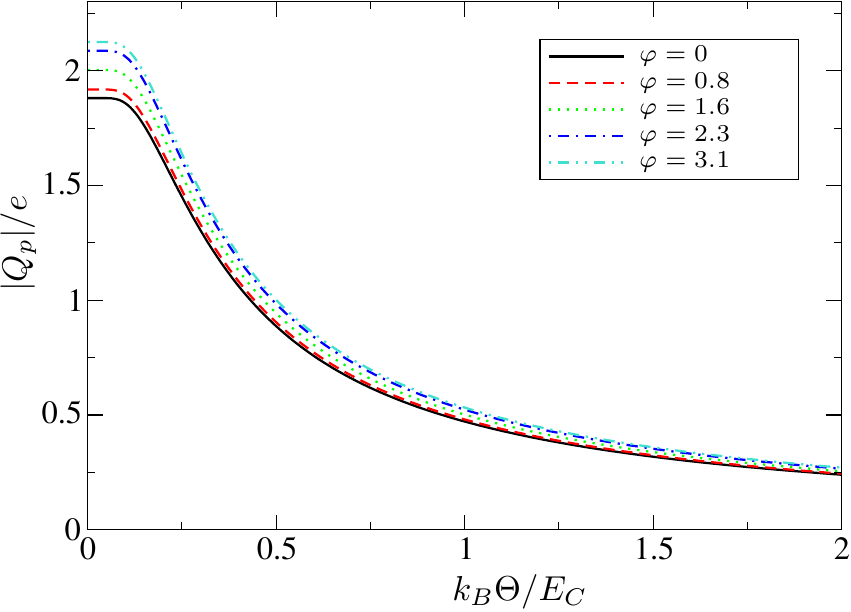}
	\end{center}
	\caption{Modulus of the pumped charge vs temperature for different values of $\varphi$. We are in the adiabatic limit ($T E_C/\hbar=8400$
	and $\Delta E_{\textrm{min}}\simeq E_C/10$). Temperature is expressed in units of $E_C/k_B$ and charge in units of $e$.}
	\label{nopump}
\end{figure}

By means of Eq.(\ref{qpump53}) the pumped charge in a dissipative case is readily obtained.  An example is shown in Fig.\ref{nopump}.

%%%%%%%%%%%%%%%%%%%%%%%%%%%%%%%%%%%%%%%%%%%%%%
%%%% F I G U R E   7 %%%%%%%%%%%%%%%%%%%%%%%%%
%%%%%%%%%%%%%%%%%%%%%%%%%%%%%%%%%%%%%%%%%%%%%%

As it should be expected, on increasing the temperature the pumped charge decreases. The pumped charge associated to the two
Floquet states is opposite for a given cycle. Therefore the progressive mixture of the two states suppresses the size of the pumping.

\subsubsection{Floquet states in the adiabatic limit}
In the limiting case of an adiabatic cycle our approach to pumping should reduce to the one discussed in Ref.~\onlinecite{pekola10}. In the
remaining of this section we will address this point by analyzing the Floquet states in the adiabatic approximation.
Floquet states are the eigenstates of the operator $ \hat{H}(t)-i\hbar\partial_t $. We assume a non-degenerate spectrum with instantaneous
eigenvectors and eigenvalues given respectively by  $\{\ket{k(t)}\}$ and $\{E_k(t)\}$. Adiabatic condition require that
$\alpha\equiv 1/(T\min\omega_{kl})\ll1 $ where  $\omega_{kl}(t)\equiv\big(E_k(t)-E_l(t)\big)/\hbar$.   In the basis of the instantaneous
eigenstates the operator to be diagonalized reads $E_k(t)\delta_{kl}-\hbar w_{kl}(t)$ where $w_{kl}(t)=\ww{k}{l}{t}$.
To first order in $\alpha$ the corresponding eigenvectors are
\begin{equation}
\label{sabbio}
	\ket{\Phi_k(t)}=\ket{k(t)}-\sum_{l\neq k}\frac{w_{lk}(t)}{\omega_{kl}(t)}\ket{l(t)}+\mathcal{O}(\alpha^2)
\end{equation}
with eigenvalues
\begin{equation}
	\label{am1}
	\epsilon_k=\frac{1}{T}\int_0^{T}\big(E_k(t)-\hbar w_{kk}(t)\Big)\ud t+\mathcal{O}(\alpha^2).
\end{equation}
It can be shown that these eigenvalues are invariant under gauge transformations.
These states are eigenstates of the operator
\begin{equation}
	\label{opb}
	\hat{D}(t)^\dagger \hH(t)\hat{D}(t)-i\hbar \hat{D}(t)^\dagger \dot{\hat{D}}(t)
\end{equation}
where $\hat{D}$ is the  the transformation from a given (time-independent) basis of the system Hamiltonian.
It was argued in Ref.~\onlinecite{pekola10} that the system relaxes in the basis of Eq.(\ref{opb}). This is what is contained in the
result of Eq.(\ref{qcicn9}) for the pumped charge.

\section{Conclusions}
\label{conclusions}

In this paper we developed a new approach to Cooper pair pumping  using Floquet theory of periodically driven quantum systems.  We found
that  the pumped charge can be expressed in a very natural way  in terms of the Floquet eigenstates  (in the adiabatic limit there is a clear
connection between Floquet exponents and states and the geometric phase acquired by the system). This approach does not require
the adiabatic limit to hold, and can be used to work out the pumped charge out of the adiabatic regime (see Fig.\ref{p21}). We further extended
to the dissipative case where we provided  a general formula to compute pumping. In order to demonstrate the power of our approach
we applied it to the case of Cooper pair sluice. In the known limits we recovered previous results. We further discussed new regimes
which are now addressable due to the expressions for the pumped charge given in Eq.(\ref{qcicnp53}).\\

\acknowledgments
We acknowledge very useful discussions with A. Shnirmann and J.P. Pekola. This work was supported by
EU through the	projects GEOMDISS, QNEMS, SOLID and NANOCTM.


\begin{thebibliography} {99}

\bibitem{thouless83}
	D.J.~Thouless, Phys. Rev. B {\bf 27}, 6083 (1983).
\bibitem{brouwer98}
        P.W.~Brouwer, Phys. Rev. B {\bf 58}, R10135 (1998).
\bibitem{kouwenhoven91}
	L.P. Kouwenhoven, A.T. Johnson, N.C. van der Vaart, C.J.P.M Harmans, and C.T. Foxon,
	Phys. Rev. Lett. {\bf 67}, 1626 (1991).
\bibitem{pothier92}
	H. Pothier, P. Lafarge, C. Urbina, D. Esteve, and M. H. Devoret, Europhys. Lett. {\bf 17}, 249 (1992).
\bibitem{pumpgeneral}
	F. Zhou, B. Spivak, and B. Altshuler, Phys. Rev. Lett. \textbf{82}, 608 (1999);
	Yu. Makhlin and A. D. Mirlin, ibid. \textbf{87}, 276803 (2001);
	O. Entin-Wohlman, A. Aharony, and Y. Levinson, Phys. Rev. B \textbf{65}, 195411 (2002);
	M. Moskalets and M. B\"{u}ttiker, Phys. Rev. B \textbf{66}, 205320 (2002);
	I. L. Aleiner and A. V. Andreev, Phys. Rev. Lett. \textbf{81}, 1286 (1998); M. Blaauboer and E. J. Heller,
	Phys. Rev. B \textbf{64}, 241301(R) (2001); B. L. Hazelzet, M. R.Wegewijs, T. H. Stoof, and Yu.
	V. Nazarov, ibid. \textbf{63}, 165313 (2001); R. Citro, N. Andrei, and Q. Niu, ibid. \textbf{68}, 165312
	(2003);  P. W. Brouwer, A. Lamacraft, and K. Flensberg, ibid. \textbf{72}, 075316 (2005); L. Arrachea,
	A. Levy Yeyati, and A. Martin-Rodero, ibid. \textbf{77}, 165326 (2008); A. R. Hern\'{a}ndez, F. A. Pinheiro,
	C. H. Lewenkopf, E. R. Mucciolo, Phys. Rev. B \textbf{80}, 115311 (2009); J. Splettstoesser, M. Governale,
	J. K\"{o}nig, and R. Fazio, Phys. Rev. Lett. \textbf{95}, 246803 (2005); E. Sela and Y. Oreg, ibid. \textbf{96}
	166802 (2006);  D. Fioretto and A. Silva, ibid. \textbf{100}, 236803 (2008).
\bibitem{geerligs91}
	L. J. Geerligs, S. M. Verbrugh, P. Hadley, J. E. Mooij, H. Pothier, P. Lafarge,
	C. Urbina, D. Esteve, and M. H. Devoret, Z. Phys. B: Condens. Matter {\bf 85}, 349 (1991).
\bibitem{pekola99}
	J. P. Pekola, J. J. Toppari, M. Aunola, M. T. Savolainen, and D. V. Averin, Phys. Rev. B {\bf 60}, 9931(1999).
\bibitem{pekola01}
	J. P. Pekola and J. J. Toppari, Phys. Rev. B {\bf 64}, 172509 (2001).
\bibitem{aunola03}
	M. Aunola and J. J. Toppari, Phys. Rev. B {\bf 68}, 020502(R) (2003).
\bibitem{fazio03}
	R. Fazio, F.W.J. Hekking, and J.P. Pekola, Phys. Rev. B {\bf 68}, 054510
	(2003).
\bibitem{niskanen03}
	A. O. Niskanen, J. P. Pekola, and H. Sepp\"a, Phys. Rev. Lett. {\bf 91}, 177003 (2003)
\bibitem{governale05}
	M. Governale, F. Taddei, R. Fazio and F. W. J. Hekking, Phys. Rev. Lett. {\bf 95}, 256801 (2005).
\bibitem{mottonen06}
	M. M\"ott\"onen, J.P. Pekola, J.J. Vartiainen, V. Brosco, and F.W. J. Hekking, Phys. Rev. B {\bf 73}, 214523 (2006).
\bibitem{leone08}
	R. Leone, L. P. Levy, and P. Lafarge, Phys. Rev. Lett. {\bf 100}, 117001 (2008).
\bibitem{brosco08}
	V. Brosco, R.Fazio, F.W.J. Hekking and A. Joye, Phys. Rev. Lett. {\bf 100}, 027002 (2008).
\bibitem{pirkkalainen10}
	J.-M. Pirkkalainen, P. Solinas, J. P. Pekola, and M. M\"{o}tt\"{o}nen, Phys. Rev. B {\bf 81}, 174506 (2010).
\bibitem{vartianen07}
 	J. J. Vartiainen, M. M{\"{o}}tt{\"{o}}nen, and J. P. Pekola, Appl. Phys. Lett. {\bf 90}, 082102 (2007).
\bibitem{mottonen08}
	M. M\"ott\"onen, J.J. Vartiainen, and J.P. Pekola, Phys. Rev. Lett. {\bf 100}, 177201 (2008).
\bibitem{avron00}
	J. E. Avron, A. Elgart, G.M. Graf, L. Sadun, Phys. Rev. B {\bf 62}, R10618 (2000).
\bibitem{zhou03}
	H.-Q. Zhou, S. Y. Cho, and R. H. McKenzie, Phys. Rev. Lett. {\bf 91}, 186803 (2003).
\bibitem{falci00}
	G. Falci, R. Fazio, G. M. Palma, J. Siewert, V. Vedral, Nature {\bf 407}, 355 (2000).
\bibitem{wang02}
	W. Xiang-bin, M. Keiji, Phys. Rev. B {\bf 65}, 172508 (2002).
\bibitem{blais03}
	A. Blais, A. M. S. Tremblay, Phys. Rev. A {\bf 67}, 012308 (2003).
\bibitem{leek07}
	P. J. Leek, J. M. Fink, A. Blais, R. Bianchetti, M. Göppl, J. M. Gambetta, D. I. Schuster, L. Frunzio,
	R. J. Schoelkopf, A. Wallraff, Science {\bf 318}, 1889 (2007).
\bibitem{jones00}
  	J. Jones, V. Vedral, A. Ekert, and G. Castagnoli,  Nature (London) {\bf 403}, 869 (2000).
\bibitem{zanardi99}
  	P. Zanardi, M. Rasetti, Phys. Lett. A {\bf 264}, 94 (1999).
\bibitem{pekola10}
	J. P. Pekola, V. Brosco, M. M{\"{o}}tt{\"{o}}nen, P. Solinas, and A. Shnirman,
	Phys. Rev. Lett. {\bf 105}, 030401 (2010).
\bibitem{solinas10}
 	P. Solinas, M. M{\"{o}}tt{\"{o}}nen, J. Salmilehto, and J. P. Pekola, arXiv:1007.2347.
\bibitem{salmilehto10}
	J. Salmilehto, P. Solinas, J. Ankerhold, and M. M{\"{o}}tt{\"{o}}nen, Phys. Rev. A {\bf 82}, 062112 (2010).
\bibitem{geodiss}
	E. Sj\"oqvist, A. K. Pati, A. Ekert, J. S. Anandan, M. Ericsson, D. K. L. Oi,
	and V. Vedral, Phys. Rev. Lett. {\bf 85}, 2845 (2000); I. Fuentes-Guridi, J. Pachos,
	S. Bose, V. Vedral, and S. Choi, Phys. Rev. A {\bf 66}, 022102 (2002);
	G. De Chiara and G. M. Palma, Phys. Rev. Lett. {\bf 91}, 090404 (2003);
	A. Carollo, I. Fuentes-Guridi, M. F. Santos, and V. Vedral, ibid. {\bf 90}, 160402 (2003);
	M. S. Sarandy and D. A. Lidar, ibid. {\bf 95}, 250503 (2005);
	R. S. Whitney, Y. Makhlin, A. Shnirman, and Y. Gefen, {\bf 94}, 070407 (2005);
	P. Thunstr\"om, J. \AA berg, and E. Sj\"oqvist, Phys. Rev. A {\bf 72}, 022328 (2005);
	G. Florio, P. Facchi, R. Fazio, V. Giovannetti, and S. Pascazio, Phys. Rev. A {\bf 73}, 022327 (2006);
	D. Parodi, M. Sassetti, P. Solinas, P. Zanardi, and N. Zangh\`i, ibid. {\bf 73}, 052304 (2006).
\bibitem{wubs06}M. Wubs, K. Saito, S. Kohler, P. H\"{a}nggi, and Y. Kayanuma, Phys. Rev. Lett. {\bf 97}, 200404 (2006).
\bibitem{moskalets02-04}
	M. Moskalets and M. B\"{u}ttiker, Phys. Rev. B {\bf 66}, 205320 (2002); ibid. \textbf{70}, 245305 (2004).
\bibitem{grifoni98}
	M. Grifoni and P. Hanggi, Phys. Rep. {\bf 304},  229 (1998).
\bibitem{guerin93}
	S. Guerin and H. R. Jauslin, Adv. Chem. Phys. {\bf 125}, 147 (2003).
\bibitem{sambe73}
	H. Sambe, Phys. Rev. A {\bf 7}, 2203 (1973).
\bibitem{gorelik01}
	L. Y. Gorelik, A. Isacsson, Y. M. Galperin, R. I. Shekhter, and M. Jonson, Nature {\bf 411}, 454 (2001).	
\bibitem{romito03}
	A. Romito, F. Plastina, and R. Fazio
	Phys. Rev. B {\bf 68}, 140502(R) (2003).
	
\end{thebibliography}
\end{document}